\documentclass[fleqn,10pt]{article}
\usepackage{natbib}
\usepackage{amsmath}
\usepackage{xcolor}
\usepackage{hyperref}
\usepackage{graphicx} 
\usepackage{caption}  

\title{Unseen Astronomy}

\begin{document}

\flushbottom
\author{James W. Trayford}
\date{}
\maketitle
\thispagestyle{empty}


The 2025 UK National Astronomy Meeting (NAM) in Durham played host to a session titled \textit{``Unseen Astronomy"}, involving a variety of astronomy researchers in diverse fields. This unique meeting focussed on a number of novel projects exploring alternatives to purely visual means of display in Astronomy, encompassing spheres of education, communication and research, and straddling both accessible and general use applications. The successful inclusion
of such a session at a major conference reflects the explosion of interest in multimodal astronomy in recent years, and hints at its transformative potential.

As one of the session organisers, I realised the potential for multimodality in astronomy – and science more broadly – from personal experience, working with data-dense physics models. As a developer on the Evolution and Assembly of GaLaxies and their Environments (EAGLE) and COLd Ism and Better Resolution (COLIBRE)  simulations, which model the evolution of the cosmic web and its constituent galaxies over billions of years, I found that the limits to understanding were not due to a lack of information. Rather, they lay in needing to develop intuition for complex data and to communicate it effectively. Experiments with sound provided new ways to interpret these data, building on visual representations and offering alternative perspectives, and pointing towards applications for new insights into data. This inspired our Ear to the Sky project - working to apply sonification more widely - and our \textit{Sonification Tools and Resources for Analysis Using Sound Synthesis} (STRAUSS) \citep{Trayford25} underpinning the \citeauthor{AU21}. 

Here, I  aim to outline and motivate the topic of multi-modal science and consider its exciting potential. I will discuss this in the context of our own work in the area, the community building being undertaken to bring together researchers considering multi-modality, and efforts to impact astronomy at large.

\subsection*{Motivating alternative modes of data representation}

For many, data visualisation is synonymous with data communication - `showing' information via graphs or imagery, without consideration of our other senses. This association is particularly strong in a field like Astronomy: an \textit{observational} science, where for most of the field's history, we could only learn about the wider Universe by measuring the light reaching us through the vacuum of space. While astronomers have long inspired the public with stunning imagery of our Universe, sole reliance on visualisation as the only sensory ``mode" of  data display can exclude people. Furthermore, this is a distorted picture of modern astronomy. Far from the origins of the field, where astronomers would meticulously study objects by eye, the vast majority of telescope data is collected digitally and processed through data pipelines, and much of it outside the narrow band of electromagnetic radiation we can actually see. The detection of gravitational waves just over a decade ago \citep{LIGO16} heralded a new age of `multi-messenger' astronomy, with the detection of oscillating spacetime distortions, inferred to originate from the coalescence of two massive black holes over a billion light-years away. This \textit{`chirp'} was detected via the LIGO instrument, and meaningfully translated to sound - allowing us to hear this  signature of the dynamic universe that surrounds us\footnote{\href{https://www.ligo.caltech.edu/video/ligo20160211v11}{\tt www.ligo.caltech.edu/video/ligo20160211v11}}. While astronomy may be emblematic of a bias towards visual modes, the dominance of visual techniques pervades all scientific and numerical fields - the presentation of graphs, equations and data handling approaches typically all tend to assume a visual interface.  

We can consider how this orthodoxy has emerged, and question if it is always justified. For example, visual presentation is closely tied to physical print media, designating text, figures and diagrams as the primary form of practical science communication for centuries as a practical way to disseminate science. Today, the proliferation of personal, digital devices render other modes of data communication (encompassing video, sound and haptic feedback) much more viable. While visual approaches may well be optimal for many tasks, there are other approaches where different senses could be more suited, or where a multi-sensory approach can combine the strengths of our different senses for a new level of understanding. 

Crucially, opening new sensory channels to access data has transformative potential to include those who are blind or have low vision (BLV). Over-reliance on visuals excludes these groups in particular, so incorporating alternative representations helps to remove barriers to science at every level. Different modalities also have advantages for those with different learning modes or sensory preferences. Moreover, new modes of data representation gives us all new perspectives on data. This has  potential to  reveal new insights or make new discoveries in complex data that we might otherwise miss. With the understanding and interpretation of data an ever more central subject of our world, even incremental improvements could be broadly applied to yield wide-scale societal benefits.

\subsection*{Exploring different modalities}
\begin{figure}[!t]
        \centering
        \includegraphics[width=\textwidth]{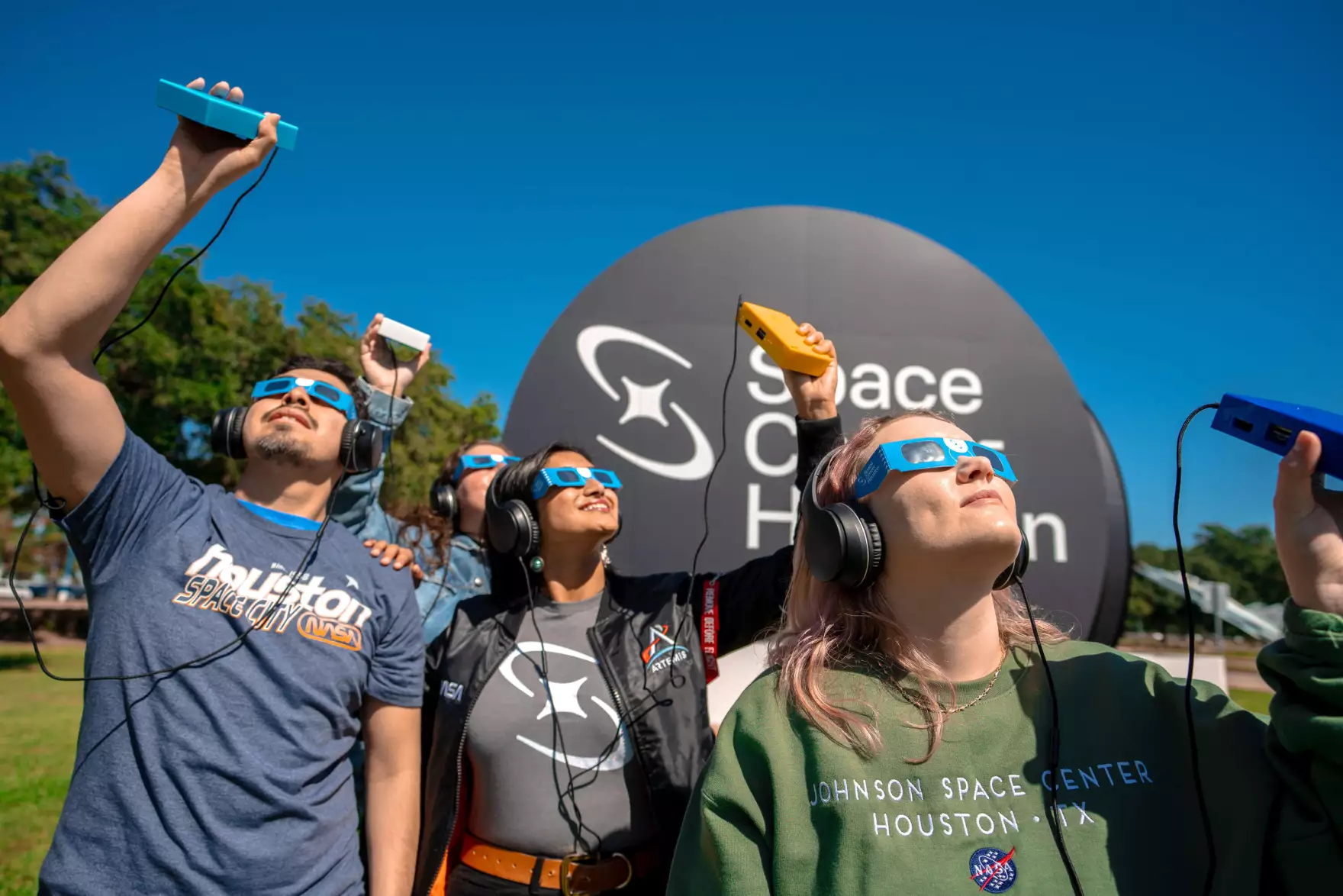}
        \caption{The \textit{Lightsound} device \citep{Bieryla20}, being used to sonify The changing light signal associated with the 2023 annular eclipse in real time. Image credit: Rochelle Pettaway.}
        \label{fig:ls}
\end{figure}

To begin exploring effective alternatives to visual representation, it is important to consider their differences. One alternative modality is using the sense of touch. Our sense of touch can be engaged through `tactile' or `haptic' stimuli, where users can explore differently contoured and textured surfaces, or interpret vibrations respectively. Touch can impart instant impressions such as \textit{`hot'}, \textit{`cold'}, \textit{`rough'} or \textit{`smooth'}, but also be used to explore complex or detailed structure through \textit{tactile graphics}. This is where imagery can be presented directly through mapping pixel brightness to contour height. Limited incorporation of tactile display for BVI accessibility is relatively established, with examples in various areas; tactile paving to alert pedestrians to roads and other hazards, embedded Braille on public interfaces and medicine packaging, or embossed relief maps. Refreshable tactile displays can be used to transfer tactile graphics digitally, though these remain relatively rare and expensive \citep[though this may be set to change, e.g.][]{Holloway24}. Smart devices that are much more widely available (such as  phones and watches) often have some limited haptic functionality. In astronomy, novel projects such as the \textit{Tactile Universe} \citep{Bonne18} focus on producing tactile graphics. 

Another mode is sound: \textit{`sonification'} is an analogous term to  \textit{`visualisation'} for communicating data using non-verbal audio. Our hearing benefits from a greater `dynamic range' in frequency than vision, perceiving up to 10 `octaves' (doublings) in sound frequency relative to the single octave of the visible spectrum (we see a range of around 400-790 Terahertz). The high sampling rate of standard audio, at around 48,000 points per second relative to the 30-60 frames per second of video, is testament to the potential for our sense of hearing to process changes over extremely fine increments in time, allowing us to detect and perceive information very rapidly. In addition, our sense of hearing is fully panoramic and always `on', making it a very useful medium for monitoring applications. Simple sound representations are  already routinely used: from the familiar variable `beep' of a parking sensor as a car approaches an obstacle or emanating from a hospital ECG monitor, to the signature notification sounds used by the different applications on our digital devices. Sound also has the advantage of being easily transmissible through broadcast or the internet, and easily reproduced by our personal phones and computers. 

While visual images and tactile surfaces can be translated quite directly, allowing multiple points in space across a two dimensional field to be perceived at once, the nature of sound requires us to develop different sonified representations of data. As a single-valued time-varying signal, using sound to represent a 2D or 3D object is a challenge. However, sound can still be perceived as `multidimensional' by making use of different phenomena manifested by the wide range of perceptible frequencies. Phenomena like the pitch, volume or rhythmic qualities of a sound, the stereoscopic balance between our left and right ears, or the higher-level musical or \textit{`timbral'}\footnote{i.e. the `character' of a sound, typically referring to the frequency harmonics or partials that constitute it.} properties can be perceived at once and allow us to encode complex information. Encoding data into the various expressive properties of sound is referred to as a the \textit{`parameter mapping'} approach, and allows us to encode rich information, but requires appropriate context and a shared sonic `language' to be meaningful. Projects such as the \textit{Audio Universe} are dedicated to exploring and developing this language. 

Other senses such as smell and taste can even be appealed to. Beyond the engaging novelty of interacting with science through these means, these different senses can have strong links to impression forming and memory \citep{Olofsson20}. This has already been applied in astronomy outreach and communication, conveying the raspberry scent of the galaxy centre or the memorable pong of comet 67P, with its potent, ammonium-rich composition \citep{Powell24}. 

Ultimately, multi-modality refers to how these different senses can be harnessed and combined to understand and communicate complex information. In a world of increasingly rich and multidimensional data, combining sensory modes also allows us to communicate more information at once, and augment our understanding. What if on top of the three dimensions we perceive visually, we could express higher dimensions simultaneously, incorporating sound or touch? With ever more complex data being generated, and its processing and even comprehension increasingly ceded to machines and AI algorithms, innovation in multi-sensory methods provides a broader toolkit to help keep human understanding and curiosity at the forefront of discovery. This is supported by research into \textit{Multisensory Learning}, where different sensory modes can combine synergistically to create a more cohesive understanding of a subject or concept, and enhance memory retention for learners \citep[e.g.][]{Shams08, Okray23}. 

\subsection*{Growing community around multisensory science}

\begin{figure}[hbt!]
        \centering
        \includegraphics[width=\textwidth]{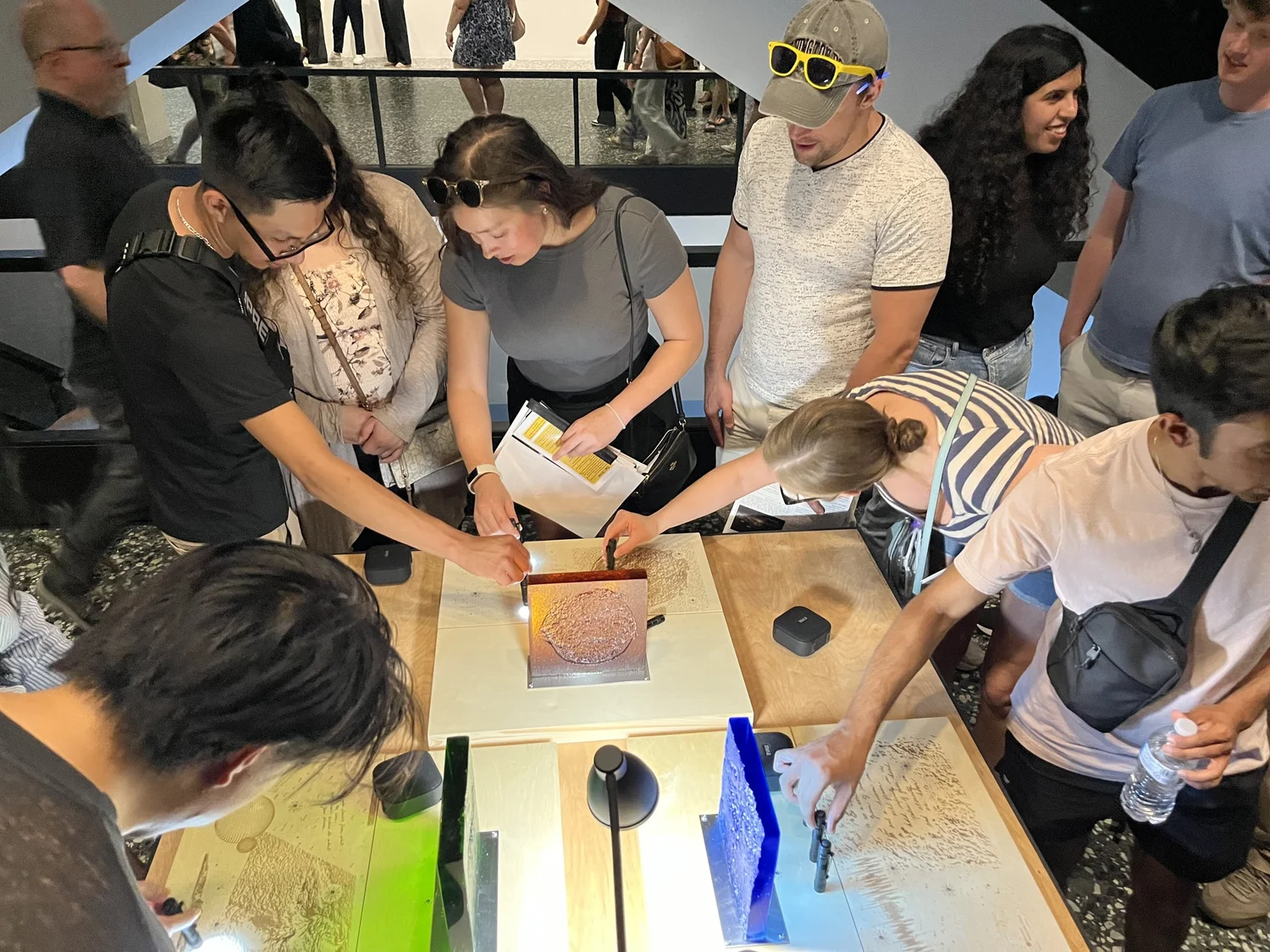}
        \caption{\textit{Infinity in our Hands}\protect\footnotemark,  exhibited during Sound Scene 2024 at the Smithsonian's Hirshhorn Museum in Washington DC. A collaboration between artists and astronomers, making use of STRAUSS as well as \textit{Chandra Observatory} models and \textit{Tactile Universe} resources. Credit: Kristine Diekman, Liz Waugh McManus, and Lisa Mansfield.} 
        \label{fig:iioh}
\end{figure}
\footnotetext{\href{https://www.lizwaughmcmanus.co.uk/infinity-in-our-hands}{\tt www.lizwaughmcmanus.co.uk/infinity-in-our-hands}}


Multisensory applications in and beyond astronomy are nothing new, with many novel examples emerging in the past decades. A particularly new aspect is the explosion in recent interest, indicated through a growth in dedicated projects   \citep[for example in sonification,][]{Zanella22}, and the impetus for a more concerted and coordinated effort to bring these techniques into standard practice to realise their promise. The \textit{Unseen Astronomy} session we held at the 2025 National Astronomer's Meeting (NAM) is an example of this, with a goal of providing a forum to foster discussion and collaboration between researchers working on multisensory applications, as well as a showcase for the broader community. This was the first session of its kind at NAM, and notable for being registered as a \textit{science} session block, acknowledging the scientific and research applications of the work alongside science communication and education.

The contributions to the session highlighted a diverse array of software and digital resources to expand the application. This included the streamlined \textit{Astronify} \citep{Brasseur24}: an open-source Python package that sonifies one-dimensional datasets, such as astronomical light-curves and spectra, using a single pitch mapping to aid in both research and data exploration. Conversely, our STRAUSS python package \citep{Trayford25} was presented as a more involved tool, allowing mappings of pitch as well as many expressive sound properties, able to produce sophisticated, multivariate sonifications, to integrate seamlessly into data analysis pipelines. This has the goal of empowering analysts to develop from sonification novices to innovators \citep{Trayford23}. In addition, a novel approach using basis function expansion to sonify galaxy imagery, exploring how spatial and structural information can be condensed into musical chords to instantly summarise complex morphological features was presented \citep{Filion25}. The \textit{Tactile Universe} project also presented the software tools for people to convert images into print-ready 3D models of tactile plates \citep{Bonne18}. 

Hardware developments, multi-sensory resources and their delivery were also a focal point, particularly regarding accessibility for the blind and low-vision (BLV) community. The \textit{LightSound} project (see Fig.~\ref{fig:ls}) \citep{Bieryla20, Davies24}, originally developed for the 2017 solar eclipse to sonify the change in brightness over the course of the event as an accessible experience, is now being updated with colour sensors for future international events. The value of multisensory engagement was further emphasised through work with the \textit{Lightyear Foundation}, sharing insights from workshops for children with complex disabilities with an immersive, multisensory approach; utilising tactile objects, scents, and kinesthesia to foster STEM engagement. The \textit{Tactile Universe} project highlighted strategic delivery, pairing its established tactile graphics with co-created 3D printed resources and sonifications regarding gravitational waves in collaboration with visually impaired students. \textit{Audio Universe} also presented the \textit{Tour of the Solar System} planetarium show \citep{Harrison22} as well as incorporation of multisensory and immersive resources with new collaborators. 

Finally, contributions also addressed analysis, both evaluating the social and educational impact of these multimodal approaches, and their efficacy in  inspecting numerical data. The results of a large-scale survey involving over 3000 participants were presented, and analysed to find that sonified NASA data yielded significant learning gains and promoted trust among both sighted and BLV audiences \citep{Arcand24}. Evalution and audience feedback for the \textit{Tour of the Solar System} planetarium show also highlighted that both sighted and BLV audiences experienced profound benefits from the sonification \citep{Harrison23}. In the sphere of data inspection, basis function expansion showed promise as a means of representing complex galaxy morphology through sound. Audience feedback and co-creation of sonifications for astrophysical light-curves was also ongoing with promising results (for example audiences able to draw graphical curves from the sounds). Finally, the use of \textit{`spectral audification'} to directly sonify astrophysical spectra, and explore unwieldy `datacubes' produced by modern telescopes, was explored with user testing indicating that users could identify key features of spectra from their sound alone \citep{Trayford23b}.

\begin{figure}[!hbt]
        \centering
        \includegraphics[width=\textwidth]{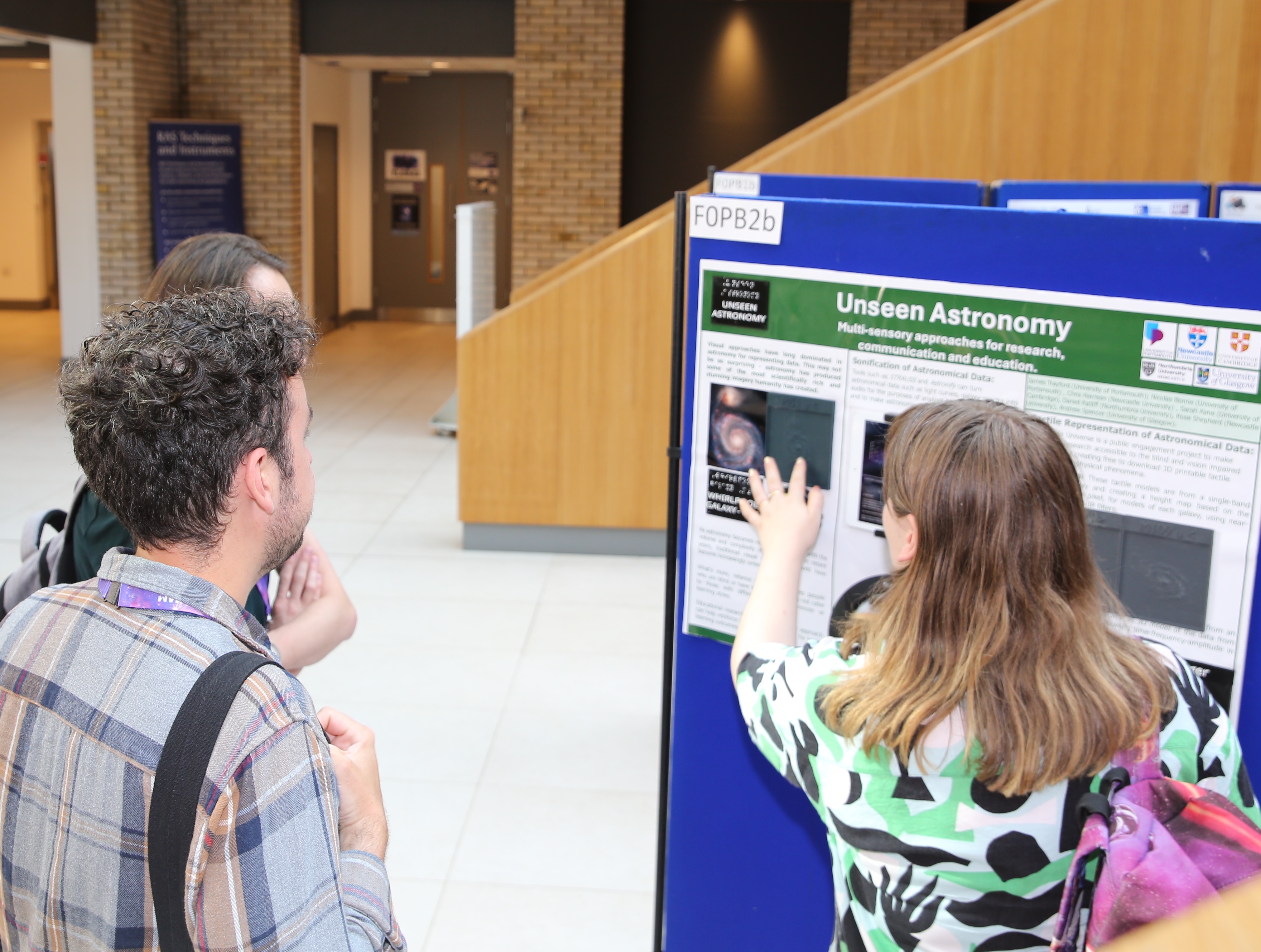}
        \caption{Participants exploring the multimodal session poster for \textit{`Unseen Astronomy'}, exploring multi-sensory approaches in astronomy at the \textit{National Astronomy Meeting}, 2025. Image credit: Durham University.} 
        \label{fig:ua}
\end{figure}

The session was accompanied by a multi-modal poster incorporating tactile graphics and audiovisual resources in the form of an accompanying video playlist, sparking a lot of general interest and discussion beyond the session itself. Figure~\ref{fig:ua} shows a few participants interacting with the session poster.


\subsection*{Building towards the Future}

\begin{figure}[!hbt]
        \centering
        \includegraphics[width=\textwidth]{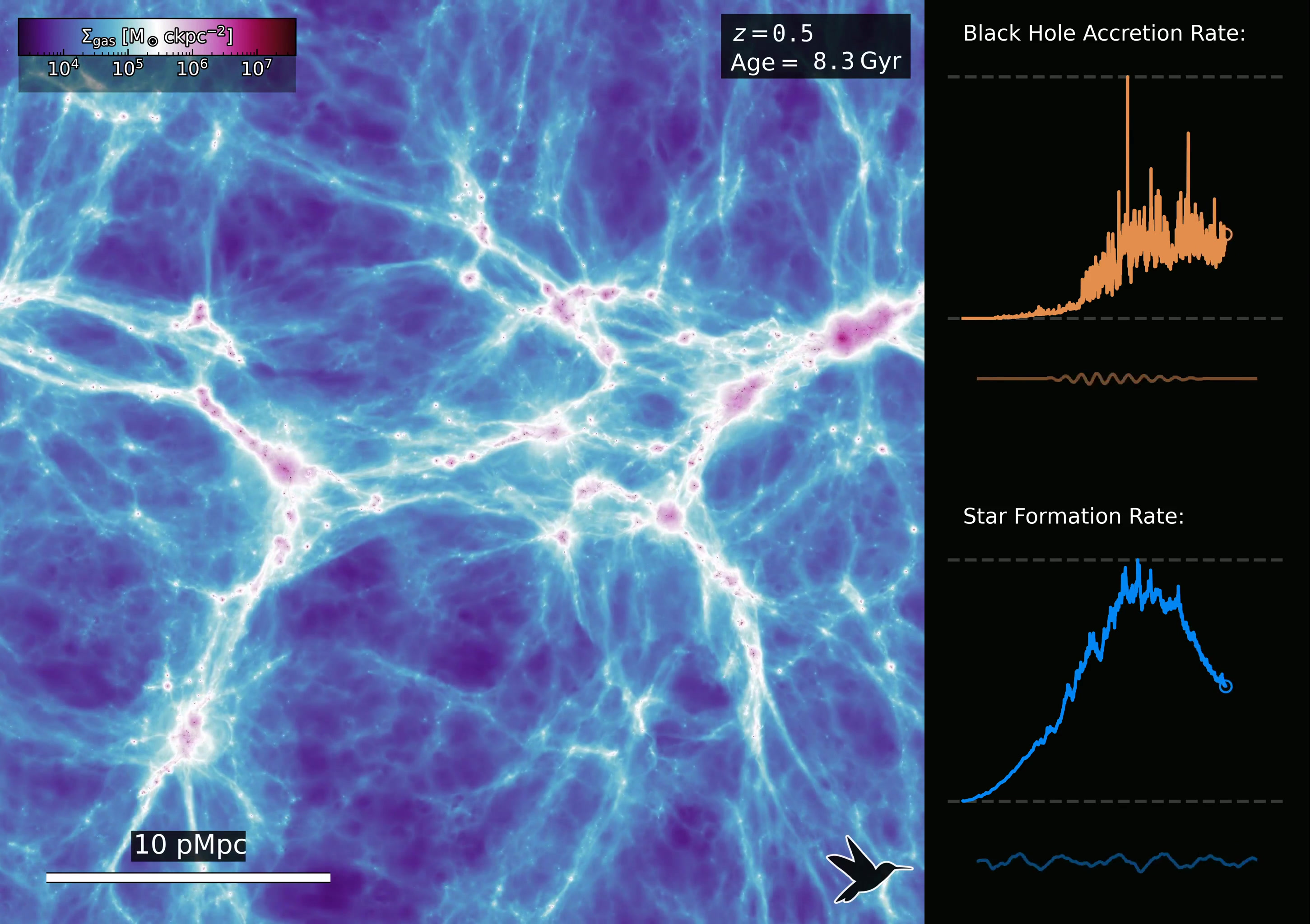}
        \caption{A static screenshot from one of the sonified videos produced for the COLIBRE simulation project, which animates the evolution of gas density on cosmic scales. This simultaneously visualises the evolution of the ‘cosmic web` (left),while sonifying key physical properties, illustrated using graphs and their accompanying waveforms (right)$^6$. Credit: The COLIBRE Team \citep{Schaye25}}
        \label{fig:colibre}
\end{figure}

While we can motivate and build communities around engaging non-visual sensory modes, sparking change across a field like astronomy at large is a considerable challenge. How we interface with data is very fundamental, and visual biases are deeply ingrained, even down to our use of language (`viewing' or `inspecting' data, `looking at the numbers', etc). This gives visual approaches a significant head-start over our other senses; just as we've built up our understanding of graphical representations of data from elementary-level education, new modes likely require a similar, long-term investment to realise their potential. What's more, institutional inertia can also act as a road-block; scientific publishers historically have tended to support only textual or graphical formats, and rely on file formats formats such as PDF which typically only reproduce information in  this form. Even scientific seminar and conference spaces  tend to put much more thought into visual display than sound: for example, while screens or projectors are almost universal, sound systems are less common and may be limited in terms of functionality.

A strategy is therefore important to realise the promise of multi-modal approaches. Alongside dedicated meetings in astronomy settings, having astronomers interface with pre-existing communities interested in different sensory modes provides a wealth of pre-existing work: for example in the area of  sonification, communities such as the \textit{International Community for Auditory Display}  or the \textit{Audio Mostly} conference. Efforts linking these communities together, such as the \textit{Audible Universe} meeting series can be particularly informative \citep{Misdariis22}. In the areas of sonification, sound design and musical expertise can be drawn from and improve the general audio quality and aesthetic appeal of sonifications, something that is particularly important for people's first interaction with the medium. The \textit{Infinity in our Hands} exhibit, is an example of collaboration between artists and astronomers, and shown in Figure~\ref{fig:iioh}. What's more, increasing interest in the deeply entangled aspects of sound and their cultural connotations, particularly with musical choices, are now being explored \citep{GarciaBenito25, Harrison25}. 

Where to focus efforts is another important question. With data interface being so fundamental, there is value in building tools and resources across a wide range of levels from elementary education, to broad science communication, to active research contexts. There is also also an important balance to be found in programs focussed on specific groups, such as the BLV community, and a broader audience. Targeted programs have already demonstrated transformative benefits for the communities involved \citep[e.g.][]{Bonne18, Harrison22, Arcand24}. However, there is a risk that without integration into the broader field, multimodal resources can only survive as long as their dedicated delivery programs. By also demonstrating benefits to a general audience, and studies that can show real advantages to these alternative modes over vision for actual data communication or discovery in scientific terms, we not only benefit science but can encourage these techniques to move towards standard practice in the field. This itself can spur more wide and enduring accessibility benefits, following the  tenets of \textit{universal design}. Aggregating and acknowledging sonification efforts across areas and fields is also important in this respect and allows a means of monitoring the growth of the field, with the establishment of the \textit{Data Sonification Archive}\footnote{\tt \href{https://sonification.design/}{https://sonification.design/}}  \citep{Lenzi21}.

Free and/or open source tools and resources and their integration into widely used toolkits are also a powerful way to encourage wider participation in multi-modal science. Encouragingly this seems to represent the majority of tools being developed \citep{Bonne18, Foran22, Guiotto23, Brasseur24, Casado24, Trayford25}, as well as many of the resources. Early-adopter communities can lead the way by continuing to use these tools to produce data representations across different data types and fields and expose them to the public. In terms of integration, sonification for virtual observatories \citep{GarciaRiber24} are emerging, such as the Rubin Observatory \textit{Skysynth}, integrated into its \textit{Skyviewer} online interface, the STRAUSS-powered datacube listener that has been built into the JDAViz data analysis software, or sonification of CALIFA datacubes making use of deep learning data reduction \citep{GarciaBenito24}.

Producing, say, sonified data products alongside visualisations for major scientific projects helps expose a broader audience to multimodal data representation, with recent examples such as the evolutionary videos of the COLIBRE cosmological simulations (Figure~\ref{fig:colibre}\footnote{\tt \href{https://colibre-simulations.org/videos}{https://https://colibre-simulations.org/videos}}).

\begin{figure}[!hbt]
        \centering
        \includegraphics[width=\textwidth]{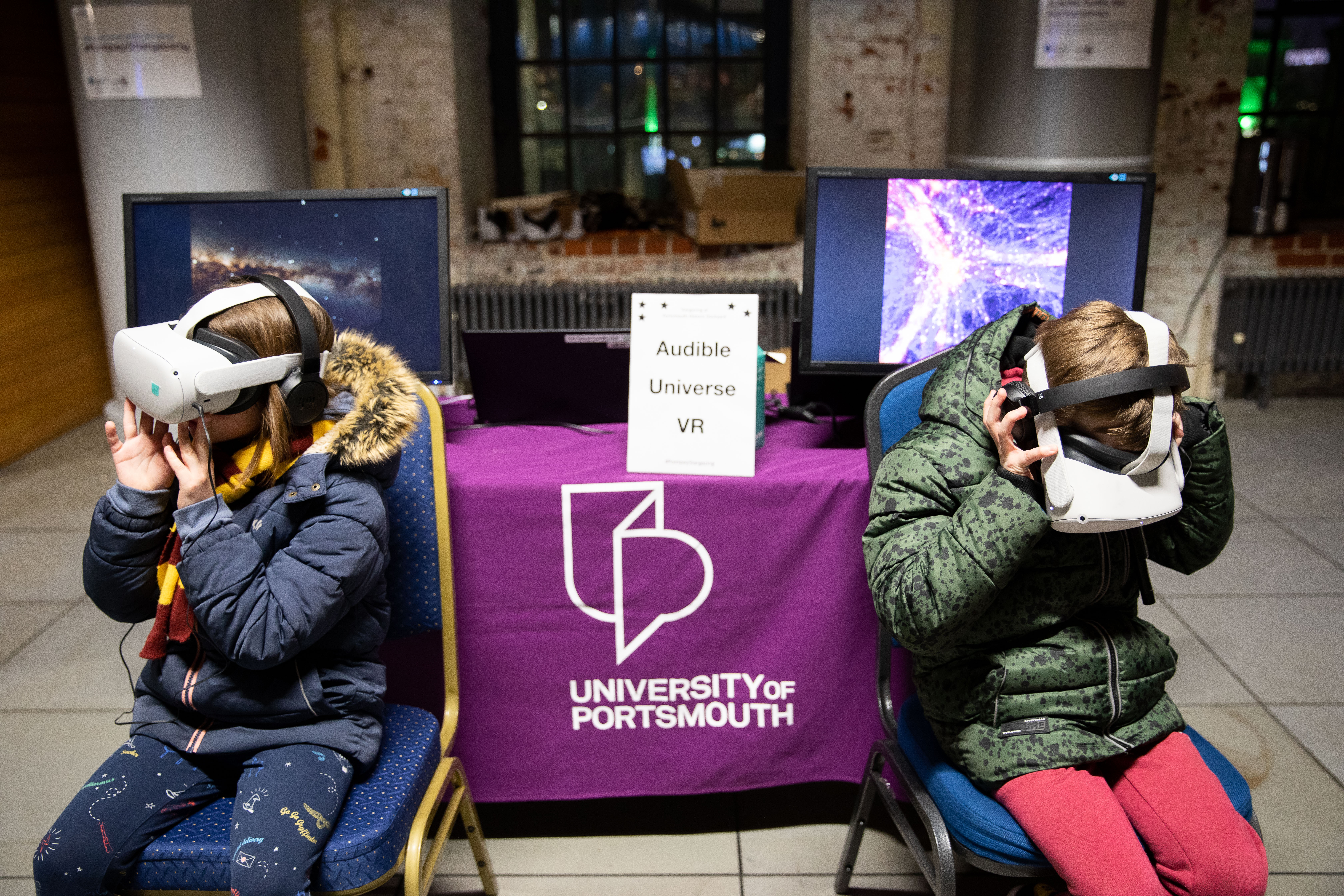}
        \caption{Immersive technology can be used to combine sound and even haptic feedback with visuals to give users a wholly new perspective. Here, children enjoy the ‘Cosmic Flythrough` and ‘Stars Appearing` sonified sequences (see {\tt \href{https://audiouniverse.org}{https://audiouniverse.org}}), with sound generated by STRAUSS. Credit: Karen Bornhoft.}
        \label{fig:vr}
\end{figure}

Finally, we can consider the place of multi-modal approaches in the rapidly changing landscapes of data and technology. The volume and rate of data collected by astronomical facilities and generated by simulations has risen exponentially in the last decades. Its also becoming increasingly complex, with \textit{`high dimensionality'} data, such as hyperspectral data cubes, or models with many hundreds of variables. AI applications can aid analysis and can be used to reduce the dimensionality of data, but further abstract humans away from the actual data we collect on our Universe. Having more channels to interpret data, or that can work together to convey more information at once, can help us maintain our understanding. With the rise of screen-less smart devices, that can be interacted with reliably through voice interface,  haptic or sound display are set to become an increasingly important tool for both a more accessible and better informed world. 

Finally, we can consider the place of multimodal approaches in the rapidly evolving landscapes of data and technology. The volume and rate of data collected by astronomical facilities and generated by simulations have risen exponentially over the
past decades. Data are also becoming increasingly complex, with ‘high-dimensional’ data sets such as hyperspectral data cubes or models comprising hundreds of variables. AI applications can aid analysis and reduce data dimensionality, but they can also further abstract humans away from the actual data we collect about our universe. Having more channels to interpret data, or combining multiple channels
to convey more information simultaneously, can
help maintain human understanding. Immersive technologies,such as virtual reality,have already proven a popular way of communicating and engaging the public in outreach contexts (Figure~\ref{fig:vr}). AI applications can aid analysis and reduce data dimensionality, but they can also further abstract the actual data we
collect about our universe away from humans.

\subsubsection*{Acknowledgments}
JT acknowledges helpful comments from Dr Chris Harrison and Rose Shepherd. JT is funded by a STFC Early Stage Research and Development Award (ST/X004651/1)












\footnotesize
\bibliographystyle{mnras}
\bibliography{sample}

\begin{figure}[hbt!]
        \includegraphics[width=0.3\textwidth]{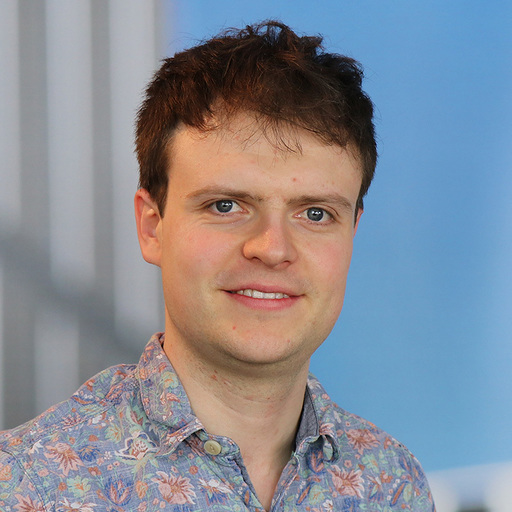}
        \caption*{\footnotesize Dr. James Trayford is an astrophysicist working as a Research Fellow at the University of Portsmouth's Institute for Cosmology \& Gravitation. Dr. Trayford has developed models of galaxy evolution, working on the recent COLIBRE simulations, and is currently leading the \textit{Ear to the Sky} project, funded by an STFC Early Stage Research and Development Award (ST/X004651/1), to develop the STRAUSS python package and sonification applications.}
\end{figure}

\end{document}